\documentclass[twocolumn,aps,pra,showpacs,showkeysfloatfix,reprint,superscriptaddress]{revtex4-1}
\usepackage{graphicx}
\usepackage[latin1]{inputenc}
\usepackage{amsmath}
\usepackage{amssymb}
\usepackage{color}

\usepackage{mathtools}
\usepackage{float}
\usepackage[caption = false]{subfig}
\usepackage{hyperref}
\usepackage[T1]{fontenc}
\usepackage{fancyhdr}
\usepackage[english]{babel}
\usepackage{bm}

\newcommand{\ket}[1]{\left| #1 \right\rangle}
\newcommand{\bra}[1]{\left\langle #1 \right|}
\newcommand{\ketbra}[2]{\left| #1 \right\rangle\hspace{-0.1cm}\left\langle #2 
\right|}

\newcommand{\nn}{\nonumber}
\newcommand{\Piop}{\hat{\Pi}}
\newcommand{\Pidop}{\hat{\Pi}^\dagger}
\newcommand{\Trace}[1]{\text{Tr}\left\lbrace #1 \right\rbrace}
\newcommand{\Pop}{\hat{P}}
\newcommand{\Pdop}{\hat{P}^\dagger}
\newcommand{\rkop}{\hat{r}_k}
\newcommand{\rkdop}{\hat{r}^\dagger_k}
\newcommand{\Bop}{\hat{\mathcal{B}}}
\newcommand{\Dop}{\hat{\mathcal{D}}}
\newcommand{\Uop}{\hat{U}}

\newcommand{\rhop}{\hat{\rho}}
\newcommand{\Hop}{\hat{H}}
\newcommand{\bop}{\hat{b}}
\newcommand{\bdop}{\hat{b}^\dagger}
\newcommand{\rop}{\hat{r}}
\newcommand{\ndg}{{\phantom{\dagger}}}
\newcommand{\dt}[1]{\frac{\text{d} #1}{\text{dt}}}
\newcommand{\dg}{\dagger}
\newcommand{\ew}[1]{\pmb{\big\langle} #1 \pmb{\big\rangle}}

\newcommand{\tp}{{t^\prime}}

\newcommand{\lk}{\left(}
\newcommand{\rk}{\right)}
\newcommand{\lsz}{\left[}
\newcommand{\rsz}{\right]}
\newcommand{\lka}{\left\{}
\newcommand{\rka}{\right\}}
\begin{document}
\author{Nikolett N\'emet} 
\email{nnem614@aucklanduni.ac.nz}
\affiliation{The Dodd-Walls Centre for Photonic and Quantum Technologies, New Zealand}
\affiliation{Department of Physics, University of Auckland, Auckland, New Zealand}
\author{Scott Parkins}
\affiliation{The Dodd-Walls Centre for Photonic and Quantum Technologies, New Zealand}
\affiliation{Department of Physics, University of Auckland, Auckland, New Zealand}
\author{Andreas Knorr}
\affiliation{Nichtlineare Optik und Quantenelektronik, Institut f\"ur 
Theoretische Physik, Technische Universit\"at Berlin, Germany}
\author{Alexander Carmele}
\email{alexander.carmele@win.tu-berlin.de}
\affiliation{Nichtlineare Optik und Quantenelektronik, Institut f\"ur 
Theoretische Physik, Technische Universit\"at Berlin, Germany}

\title{Stabilizing quantum coherence against pure dephasing in 
the presence of quantum coherent feedback at finite temperature 
}
\begin{abstract}
{A general formalism to describe the dynamics of quantum emitters in
structured reservoirs is introduced. 
As an application, we investigate the optical coherence of an atom-like
emitter diagonally coupled via a link-boson to a structured bosonic 
reservoir and obtain unconventional dephasing 
dynamics due to non-Markovian quantum coherent feedback for different temperatures. 
For a two-level emitter embedded in a phonon cavity, 
preservation of finite coherence is predicted up to room temperature.}
\end{abstract}
\maketitle
\date{\today}
\selectlanguage{english}

\section{Introduction}
Few-level quantum systems such as atoms, molecules, and their artificial 
counterparts in solid state structures are envisioned as fundamental building 
blocks of quantum communication, encryption and computation 
\cite{Nielsen2010,Gardiner2015,Scully1997,Zoller2005}. 
Quantum dots in solid state systems are approximated as few-level, or in the ideal case, two-level emitters (TLE), which is similar to the way atoms are handled in quantum optics \cite{walls2008}. The interaction of these dots with the collective excitations of the surrounding bulk material - especially acoustic 
phonons - are perceived as the main threat to coherence in these systems. These decoherence processes also hinder the 
efficient functioning of quantum information processing on such platforms
\cite{May2008,Bimberg2008,Nazir2015,Iles2017}.
Optical excitation of these quantum dots restructures the local 
electron configuration, and induces lattice vibration around the equilibrium positions of the 
atomic cores in the bulk. 
While the system equilibrates, the generated phonons induce a 
wide range of phenomena such as the damping of Rabi oscillations 
\cite{Forstner2003a,Forstner2003b}, cavity-feeding \cite{Roy-Choudhury2015},
striking photon-phonon coupling \cite{Roy2011,Harsij2012}, incoherent 
excitation of the emitter \cite{Hughes2013,Weiler2012,Wigger2014}, and 
phonon-assisted quantum interferences \cite{Carmele2013,Wilson2002}. 
Thus, in comparison to isolated atomic systems,
solid state systems are generally considered to have 
larger dephasing and damping rates, most of which could be overcome only by cooling the system
to very low temperatures. However, phonon emission is still a source of decoherence even at 0~K. 
Due to the interaction with the solid state environment (via, e.g., acoustic phonons), these 
processes cannot be completely eliminated, but the idea of controlling them
has motivated theoretical investigations to turn this drawback into a useful 
and essential feature. 
In recent proposals a dissipative interaction is 
used to induce phonon-lasing \cite{Droenner2017}, ground-state cooling 
\cite{Martin2004,Jaehne2008}, or even stabilization of the dynamics 
\cite{Bounouar2015,Carmele2013,Glassl2012} 
in the strong coupling regime. 
Nevertheless, for efficient reservoir engineering, a more 
general theoretical framework is needed to understand, describe, and control a wide variety 
of different systems and structured reservoirs at finite temperatures.

\begin{figure}[b!]
\centering
\vglue -.4cm
\includegraphics[width=8.3cm]{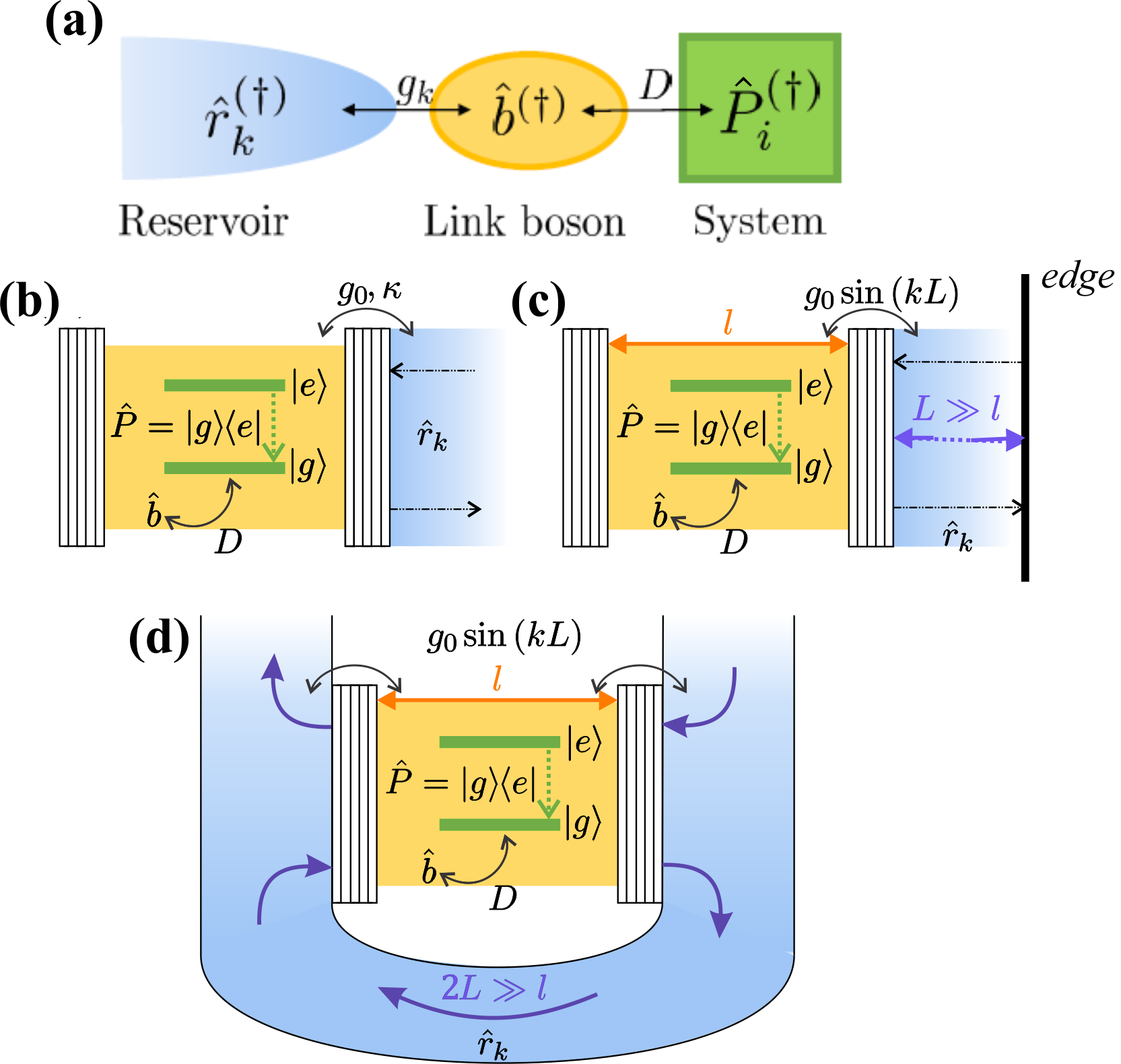}
\vglue -.2cm
\caption{(a) Schematics of potential setups to consider. An example where this approach can be applied is a quantum emitter coupled to a single-mode phonon cavity, which is embedded in a waveguide 
without edge (b), and with a perfectly reflecting edge (c). An alternative, unidirectional realization for the feedback case with chiral coupling to the waveguide can also be considered, where $2L$ is the length of the feedback loop (d).}
\label{fig:setup}
\end{figure}

In this Paper, we propose a model where a two-level emitter (TLE) is 
coupled to a dissipative boson, such as an optical phonon 
\cite{Hameau1999,Krauss1997,Carmele2010} or an 
acoustic cavity mode 
\cite{Fainstein2013,Lanzillotti-Kimura2015,Trigo2002,Kabuss2012}.
In practice, such a dissipative, single bosonic mode can be 
realized by phonon-confinement in layered structures 
\cite{Trigo2002,Winter2005,Soykal2011,Anguiano2017} or photonic-phononic 
crystals \cite{Fainstein2013,Lanzillotti-Kimura2015}. 
We address this boson as link boson since it provides a link between the system and the bosonic reservoir.
The dynamics of the link boson can be controlled via its initial state and 
interaction with the reservoir (Fig.~\ref{fig:setup}(a)).  This reservoir of the link boson (LB) has typically infinite degrees of freedom but can be spectrally unstructured, as in Fig.~\ref{fig:setup}(b), or structured by, e.g., a perfectly reflecting edge (Fig.~\ref{fig:setup}(c)) or an imposed chiral coupling (Fig.~\ref{fig:setup}(d)).
A reservoir-structuring as shown in Fig.~\ref{fig:setup}(c) can be realized by phononic Bragg
mirrors 
\cite{Trigo2002,Winter2005,Soykal2011,Anguiano2017,Fainstein2013,
Lanzillotti-Kimura2015}.
This way, the infinite degrees of freedom of the solid state reservoir amount either 
to enhanced or suppressed dephasing of the TLE 
\cite{Pirkkalainen2013,Carmele2013} and may even
introduce, e.g., time-delayed feedback for the system's dynamics \cite{Faulstich2017}.

The other side of the coin is that there are unique and yet not understood non-Markovian effects present in solid state systems intrinsically 
\cite{Cassabois2008,Moody2012,Thoma2016,Liu2018}.  The introduction of a link boson - with its extra degrees of freedom - facilitates the development of an effective model of these non-trivial dephasing dynamics. One important feature of our introduced method in that respect is the ability to separately control the Markovian ($g_0$) and non-Markovian ($D$ and $L$) characteristics.
%
%

%

The interaction between acoustic phonons and the TLE can be formulated as a 
level-shifting, pure-dephasing-type interaction \cite{Bimberg2008,Nazir2015}, where an additional Ohmic spectral density is considered for the continuum of 
phonon modes that introduces
non-Markovian correlations
\cite{Leggett1987,Besombes2001,Favero2003}.
This model, the independent boson model,
accurately describes the dephasing dynamics of the emitter in the linear regime. However, due to the absence of excitation exchange, it is limited to unilateral phase-destroying processes 
without the build-up of quantum entanglement between emitter
and reservoir
\cite{Petruccione2007,Nazir2015,Krummheuer2002,Pazy2002,May2008,Agarwal2012}.
Here, by connecting 
the emitter via a LB to the solid state reservoir and allowing for excitation 
transfer between them, we demonstrate how one can overcome this limitation.
%
The considered exchange of excitations introduces quantum interference into
the system dynamics, enables reservoir-engineered 
quantum coherence, and forms a robust recipe for quantum state preparation
\cite{Ramos2013,Law1996}.
Focusing only on boson-emitter interactions, the investigated model in the present work can be applied in diverse settings, ranging from acoustic phonons and 
quantum dots to a wide range of other setups, especially in the 
context of circuit QED. 
Micro- and nanomechanical resonators coupled to Nitrogen-vacancy centres 
\cite{Rabl2009,Arcizet2011}, Cooper boxes 
\cite{Armour2002,Irish2003,Martin2004,Jaehne2008,LaHaye2009,Stadler2017}, 
Josephson junctions and SQUIDs \cite{Xue2007,Etaki2008,Khosla2017} are perfect 
platforms for testing fundamental quantum mechanics as well as quantum 
non-demolition measurement of the TLE 
\cite{Johnson2010,Reiserer2013}. 
Also, in optical lattices, recent experiments show strong control over single 
vibrational modes in ion traps with generalized pure dephasing coupling 
\cite{Lemmer2017}.
A straightforward extension of these systems are when the quantum 
emitter is located on the surface of the mechanical resonator 
interacting with the vibrational mode via a strain-mediated coupling 
\cite{Wilson-Rae2004,Yeo2013,Munsch2017,Ramos2013}.
In the following, we demonstrate that the introduction of a link boson
allows a simple yet powerful analytical solution for a wide range of dynamics
unattainable within the typical independent boson model 
\cite{Krummheuer2002,May2008,Agarwal2012}. The presented model demonstrates the link between the two kinds of non-Markovianity, i.e., the one that is observed as naturally appearing in solid state systems and the one that is the topical case of artificially structured reservoirs in quantum optics. 
An example of such a structured reservoir that we consider is a coherent, feedback-type,
reservoir-LB interaction, which can be used to steer the dynamics of the TLE with the goal of coherence preservation.
Moreover, even though there are numerical methods that in principle are capable of characterizing the effect of  quantum coherent feedback at a finite temperature \cite{Pichler2016,Whalen2016}, to our knowledge, this is the first work to report on the exact influence of this effect on the amount of recovered coherence.

\section{Theoretical Model}\label{sec:model}
For generality, our model considers an arbitrary set of system operators 
$\hat P^{(\dg)}_i$ that can be coupled to a single bosonic mode
with annihilation (creation) operator $\hat b^{(\dg)}$.
This link boson (LB) couples the system (S)
to another bosonic reservoir described by operators $\hat r_k^{(\dg)}$; 
see Fig.\ref{fig:setup}(a). 
Thus, in its most general form, the model assumes the following Hamiltonian,
\begin{align}
\label{eq:H}
\Hop =&\ \Hop_S+\Hop_R+\Hop_{LB}\hspace{-.1cm}\lk  \hat 
b,\hat b^\dg,\hat P^\ndg_i,\hat P_i^\dg\rk ,
\end{align}
where $\Hop_S$ describes the free evolution of the system 
operators $ \Pdop_i $, i.e., $\Hop_S = \sum_i \hbar\omega_i \Pdop_i \Pop_i$.
The LB interacts with the bosonic reservoir via $ \Hop_R $,
\begin{align}
\label{eq:H_R}
\Hop_R/\hbar=& 
\omega_0 \bdop\bop\ +  
\int \left[  \omega_k \rkdop\rkop  + 
g_k(\rkdop\bop+\bdop\rkop) \right]dk,
\end{align}
where $k$ labels the lowering (raising) operator for a reservoir 
mode with wave number $k$, and $g_k$ describes the $k$-dependent interaction 
strength between the two bosonic fields.
The system and the LB interact via the interaction
Hamiltonian $\Hop_{LB}\hspace{-.1cm}\lk \hat b,\hat b^\dg,\hat 
P^\ndg_i,\hat P_i^\dg \rk$ with coupling strength $D_i$ (see~Fig.\ref{fig:setup}(a)), which will be specified
later.

The effective action of the reservoir 
$\lbrace \hat r_k^\ndg, \hat r_k^\dg \rbrace $ on
the LB is given by the solution of the LB-reservoir dynamics, as described by $\hat H_R$ 
in Eq.~\eqref{eq:H_R}. 
Using Heisenberg equations of motion, the 
LB-dynamics can be represented with a linear map,
\begin{align}
\label{eq:b_def}
\bop(t) &= F(t)\bop(0)+\int G_k(t)\rkop(0) dk,
\end{align}	
where $ F(t)$ and $G_k(t)$ are $c$-number functions satisfying
$[\hat b(t),\hat b^\dg(t)]=1$ at all times.
We provide in the following two examples for the LB-reservoir
interaction. 
Case (i): a dissipative, Markovian, irreversible 
interaction of the LB with its reservoir $ (g_k = g_0)$.
A typical example is given in Fig.~\ref{fig:setup}(b), where a 
single two-level system couples to a single mode of a 
photonic/phononic nanocavity $b^{(\dagger)}$.
The Markovian interaction leads to a 
dissipative link boson dynamics and to the propagators
\begin{align}
\label{eq:FMark}
F_{d}(t)&=e^{-(i\omega_0+\kappa)t},\\
\label{eq:GMark}
G^d_{k}(t)&=-ig_0 
\int_0^t e^{-(i\omega_0+\kappa)(t-\tp)-i\omega_k\tp}d\tp,
\end{align}
where $ \kappa=\pi g_0^2/(2c)$ is the decay rate of the LB into the continuum of 
modes of the reservoir. 
This case is well documented in the literature 
\cite{Scully1997,Gardiner2015,May2008,Weiss2012,
Huelga2012}.
On the other hand, in case (ii) we consider a $k$-dependent 
coupling $\lk g_k\rk$ structuring the reservoir mode. This is realized by a half-open 
cavity scheme for the link boson and a distant reflector in the environment, as shown in Fig.~\ref{fig:setup}(c). 
This reservoir coupling produces a coherent, time-delayed feedback to the LB $ (g_k = g_0\sin[ck\tau/2] )$, where $ 
\tau=2L/c$ is the roundtrip time and $c$ is determined by the 
dispersion relation valid in the reservoir 
\cite{Weiss2012,Breuer2002}. An equivalent representation of this feedback scheme is shown in Fig.~\ref{fig:setup}(d). In this case a chiral coupling to the waveguide imposes unidirectional flow of information back to the system. The roundtrip time now is translated into the length of the feedback loop, $2L$.

Solving exactly the reservoir-LB dynamics leads to the following 
c-number functions $F,G$
(see Appendix \ref{sec:FG} for a detailed derivation),
\begin{align}
\label{eq:Ffb}
F_{fb}(t) &= \sum_{m=0}^\infty\frac{\kappa^m}{m!}e^{-B(t-m\tau)} 
(t-m\tau)^m\theta(t-m\tau),\\
\label{eq:Gfb}
G_k^{fb}(t)&=\sum_{m=0}^\infty\frac{-i g_k 
\kappa^{m}\theta(t-m\tau)}{(A_k+B)^{m+1}}
\lsz e^{A_k(t-m\tau)}-\right.\\
&\quad-\left.e^{-B(t-m\tau)}\sum_{n=0}^m\frac{1}{n!}(A_k+B)^n 
(t-m\tau)^n\rsz,\nn
\end{align}
where $\kappa$ is defined as before
and the coefficients as $ A_k=-i\omega_k$ and $B=i\omega_0+\kappa$.
The expressions above include non-trivial feedback quantities depending
on the previous roundtrips (integer multiples of $\tau$) of the fields $ 
b$ and $ r_k$.
These terms enable some interesting quantum interference phenomena, such as 
coherence stabilization, and give 
access to the reservoir dynamics 
\cite{Kabuss2015,Dorner2002,Kabuss2016,Guimond2017}. 
\subsection{General interaction}
Let us consider an interaction Hamiltonian, $\Hop_{LB}$, involving arbitrary system and link boson operators. The main objective of the applied method is to calculate the time-evolution of a given system operator $\Pop$ analytically for different initial conditions of both the system and the reservoir, as well as for different couplings between the two.

Since the effect of the surrounding reservoir dynamics is fully
incorporated in the LB's time trace, we can evaluate the 
Liouville-von Neumann equation explicitly in the interaction picture,
\begin{align*}
-i\hbar\frac{d}{dt}\hat{\rho}_I(t)=[\Hop_{LB}(t),\hat{\rho}_I(t)],
\end{align*}
and from Eq.(\ref{eq:H},\ref{eq:H_R}),
\begin{align*}
\Hop_{LB}(t)&=\Uop(t)\Hop\Uop^\dg(t) -i\hbar \Uop(t) \dot \Uop^\dg(t),\\
\Uop(t)&=\exp{\lk i(\Hop_S+\Hop_R)t\rk},
\end{align*}
since the system operators commute with the reservoir.
Thus note, the system-LB interaction becomes time-dependent, 
and in the time-dependence the full reservoir interaction is present (see the derivation in Appendix~\ref{sec:FG}).

For a quadratic system Hamiltonian $\Hop_S$ with only real eigenvalues, there can be found a set of system operators $\{\hat{\Pi}_i\}$, the normal modes of the system, for which the time-dependence has the following form,
\begin{align}
\label{eq:Pi_evol}
\Piop_i(t)=e^{-i\omega_it}\Piop_i(0).
\end{align}
Any given system operator $\Pop$ can be constructed from these operators as
\begin{align}
\label{eq:P_expand}
\Pop(t)=\sum_i\lsz\alpha_i\Piop_i(t)+\beta_i\Pidop_i(t)\rsz,\\
\sum_i\lk|\alpha_i|^2+|\beta_i|^2\rk=1.\nn
\end{align}
If we want to interpret the normal modes as quasiparticles with bosonic or fermionic commutation relations, the conditions
\begin{align*}
|\alpha_i|^2\pm|\beta_i|^2=1
\end{align*}
also apply, with "$-$" corresponding to bosons and "$+$" to fermions.

Using the Liouville-von Neumann equation, for each of the normal mode operators we can prescribe the following equation
\begin{align}
\frac{d\hat{\rho}_I}{dt}\Piop_i(t)=-\frac{i}{\hbar}\lsz\Hop_{LB},\rhop_I\rsz\Piop_i(t).
\end{align}
If the commutation relationship between the interaction Hamiltonian and $\Piop_i(0)$ can be written as
\begin{align}
\label{eq:H_comm}
\lsz\Hop_{LB}(t),\Piop_i(0)\rsz=\hat{C}_i(t),
\end{align}
and using the time-evolution of $\Piop_i$, Eq.~(\ref{eq:Pi_evol}), the following equation of motion can be obtained,
\begin{align}
\text{Tr}\lk\frac{d\rhop_{\Pi_i}(t)}{dt}\rk&=-\frac{i}{\hbar}Tr\lka\lsz\Hop_{LB},\rhop_{\Pi_i}(t)\rsz-\rhop_I\hat{C}_i(t)\rka=\nn\\
&=\frac{i}{\hbar}Tr\lka\lsz\Piop_i^{-1}(0)\hat{C}_i(t)\rsz\rhop_{\Pi_i}(t)\rka\\
\rhop_{\Pi_i}(t)&=\rhop_I(t)\Piop_i(0),
\end{align}
which, due to the linear property of the trace, translates into
\begin{align}
\frac{d}{dt}\ew{\Piop_i(t)}&=-\frac{i}{\hbar}\ew{\Dop_i(t)\Piop_i(t)},\\
\Dop_i(t) &= -\Piop_i^{-1}(0)\hat{C}_i(t).
\end{align}
A stroboscopic solution for this equation for short enough time steps gives
\begin{align*}
\ew{\Piop_i(t+\Delta t)}&=\left\langle\lk1-\frac{i}{\hbar}\Dop_i(t)\Delta t\rk\Piop_i(t)\right\rangle\approx\nn\\
&\approx \left\langle e^{-\frac{i}{\hbar}\Dop_i(t)\Delta t}e^{-\frac{i}{\hbar}\Dop_i(t-\Delta t)\Delta t}\Piop_i(t-\Delta t)\right\rangle,
\end{align*}
which, by applying the Baker-Campbell-Hausdorff formula, turns into
\begin{align}
\label{eq:Pi_exp}
&\ew{\Piop_i(t)}=\left\langle\exp{\lsz-\frac{i}{\hbar}\int_0^t\Dop_i(t_1)dt_1\rsz}\cdot\right.\\
&\hspace{.6cm}\cdot\left.\exp{\lsz-\frac{1}{2\hbar^2}\int_0^t\int_0^{t_1}\lsz\Dop_i(t_1),\Dop_i(t_2)\rsz dt_2dt_1\rsz}\Piop_i(0)\right\rangle.\nn
\end{align}
The time evolution of the expectation value of the original system operator $\Pop(t)$ can be constructed from these expectation values by using (\ref{eq:P_expand}).
\subsection{TLE-phonon interaction}
As we saw in the previous subsection, the 
model is applicable to a wide range of system-LB interactions. To provide
an example, we consider the interaction between an optically excited spin/two-level emitter and a
single-phonon mode \cite{Feynman2000}.
This specific example (see~Figs.~\ref{fig:setup}(b-d)) is of great importance, as it 
describes the loss of coherence in quantum systems enforced by level-shifts 
of the emitter \cite{Calarco2003,Borri2005,Krugel2005} or 
between quantum emission processes influencing the 
indistinguishability of photons
\cite{Thoma2016,Kaer2013,Unsleber2015},
\begin{align}
\label{eq:HI_def}
\Hop_{LB}(t) = \hbar D[\bop(t)+\bdop(t)]\Pdop(t)\Pop(t),
\end{align}
where $\Pop=\ketbra{g}{e}$ is the lowering operator of the TLE.
Note, that this model implies the pure dephasing limit, where the 
level-spacing of the 
TLE $\lk\hbar\omega_i=\hbar\omega_{eg}\rk$ is much greater than the energy of the 
LB $\lk\hbar\omega_0\rk$, and the population for the TLE, $\ew{\hat P^\dg 
\hat P}$, is not influenced by the reservoir. 
In contrast to the population, however, the coherence amplitude 
$\ew{{\hat P}^{(\dg)}}$ is strongly affected, showing, for instance, non-trivial dynamics due to unconventional coupling between the system and reservoir.

A relevant effect to investigate is the controllability of the coherence 
$\ew{{\hat P}^{(\dg)}}$ by the reservoir coupling, $ g_k $, 
and the description of different, unconventional dephasing dynamics. 
We choose the coherence as our observable, $\ew{\Pop(t)}=\Trace{\hat \rho(t) \hat P(t)}$, and
define the absolute square of the polarization 
at time $ t $ normalized by its initial value as our
figure of merit, i.e.,
\begin{align}
\label{eq:eta}
\eta(t)= \frac{|\ew{\hat P(t)}|^2}{|\ew{\hat P(0)}|^2}.
\end{align}

The goal of the modification of the LB-reservoir dynamics, 
$ g_k : g_0 \rightarrow g_0 \sin(kL)$ 
\cite{Kabuss2016,Faulstich2017,Dorner2002,
Guimond2017,Grimsmo2014}, is to restore/stabilize as much of the initial 
coherence for as long as possible.
As there is no direct driving considered for the TLS, we have
\begin{align}
\label{eq:P_evol}
\Pop(t)&=\Pop(0)e^{-i\omega_{eg}t},
\end{align}
which is a special case of the situation described in the previous section. Here, $\Piop(0)=\Pop(0)$, and the commutator (\ref{eq:H_comm}) has the following form
\begin{align*}
\lsz\Hop_{LB},\Pop(0)\rsz=-D[\bop(t)+\bdop(t)]\Pop(t)\Pdop(t)\Pop(t).
\end{align*}
Thus equation (\ref{eq:Pi_exp}) turns into
\begin{align}
\hat \rho^I_{\Pop}(t) =& 
\exp
\left[
-i
\int_0^t 
\Hop_{LB}(t_1) 
dt_1
\right] \cdot\\ & \nonumber
\exp
\left[
-\frac{1}{2}
\int_0^t 
\int_0^{t_1} [\Hop_{LB}(t_1),\Hop_{LB}(t_2)]dt_2
dt_1
\right]
\hat
\rho^I_{\Pop}(0),
\end{align}
which can also be obtained by a similar but more direct method used in Appendix \ref{sec:sp-LB}.

As for the LB-operators, the following commutation 
relation is valid: $ [\Bop(t_1),[\Bop(t_1),\Bop(t_2)]]=0$ (where $\Bop(t)=\bop(t)+\bdop(t)$), and thus,
\begin{widetext}
\begin{align}
\label{eq:20}
\rhop_P(t)=&\ \exp\lka\lk-i\int_0^t \Bop(t_1)dt_1-\frac{1}{2} \int_0^t 
\int_0^{t_1}[\Bop(t_1),\Bop(t_2)]dt_2dt_1\rk\Pop^\dg(0) \Pop(0)\rka\rhop_P(0) .
\end{align}
\end{widetext}

Taking the trace of the expression above gives the expected time-evolution of the coherence, which can be written as
\begin{align}
\Trace{\rhop_P(t)} =&\sum_{m,i,n_k}\bra{i,m,\lbrace 
n_k\rbrace}\rhop_P(t)\ket{i,m,\lbrace n_k \rbrace},
\end{align}
where $ i=e,g $ accounts for the electronic states, $ m=0...N $ counts the 
phonon number in the cavity, and $ n_k $ refers to the reservoir states.

Substituting equation (\ref{eq:20}) in, and collecting together the link-boson part as
\begin{align*}
\hat\Upsilon(t) = -i\int_0^t \Bop(t_1)dt_1-\frac{1}{2} \int_0^t 
\int_0^{t_1}[\Bop(t_1),\Bop(t_2)]dt_2dt_1,
\end{align*}
we obtain
\begin{widetext}
\begin{align}
&\Trace{\rhop_P(t)} =\sum_{m,i,n_k}\bra{i,m,\lbrace n_k \rbrace}e^{\hat\Upsilon(t)\Pop^\dg(0) \Pop(0) 
}\rhop_I(0)\Pop(0)\ket{i,m,\lbrace n_k \rbrace}.\nn
\end{align}
\end{widetext}
Considering the electronic part of this expression, only the 
expectation value taken with the excited state gives a contribution, as $ 
[\Upsilon(t),\Pop^\dg(0)\Pop(0)]=0$ and $ (\Pop^\dg \Pop)^n=\Pop^\dg \Pop $, so
\begin{align}
\bra{e} e^{\Upsilon(t) \Pop^\dg(0)\Pop(0)}=e^{\Upsilon(t)}\bra{e}.\end{align}
The final expression for the general solution can be written as 
\begin{align}
\label{eq:rhop_P}
\rhop^I_{\hat P}(t) 
 =&\ \exp{\lk-i\lka\omega_{eg}t D\lsz \gamma(t) 
\bop(0)+\gamma^*(t)\bdop(0)\rsz+ \right.\right.}\nn\\
 &{\quad\qquad\left.\left. +D\int \lsz N_k(t) \rkop(0)+N_k^*(t)\rkdop(0)\rsz 
dk+\right.\right.}\nn\\
&{\quad\qquad\left.\left.+D^2\phi(t)\rka\Pop^\dg \Pop \rk\rhop^I_{\hat P}(0)} ,
\end{align}
with 
\begin{align}
\label{eq:gam_def}
\gamma(t) =& \int_0^t F(t^\prime)dt^\prime, \qquad
N_k(t) = \int_0^t G_k(\tp)d\tp,\\
\phi(t)=&
\text{Im}\left[
\int_0^t (F(\tp)\gamma^*(\tp)
+\int G_k(\tp)N_k^*(\tp) dk) d\tp
\right]. \nonumber
\end{align}
Such an analytic expression can be
derived for an arbitrary mixed state
and temperature-dependent reservoir states, as well as for 
coherent feedback with structured reservoirs at a finite
temperature. 
The trace of the above expression with a given set of initial 
conditions for the system, the link boson, and the reservoir, provides the following expectation 
value of the time-dependent coherence,
\begin{align}
\label{eq:sigma_def}
& 
\ew{\hat 
P(t)}=\Trace{\rhop^I_{\hat 
P}(t)}=\sigma_S(t)\sigma_{LB}(t)\sigma_R(t), \\
&\sigma_S(t)=\bra{e}\rhop_S(0)\ket{g}=\ew{P(0)} e^{-i\omega_{eg}t-i\phi(t)}, \label{eq:sigmas}\\
&\sigma_{LB}(t)=\sum_{m}\lka\bra{m}e^{-i \lsz\gamma(t) 
\bop(0)+\gamma^*(t)\bop^\dg(0)\rsz}\rhop_b(0)\ket{m}\rka,\nn\\
& \nonumber \sigma_R(t)
=\sum_{ \lbrace n_k \rbrace } \bra{ \lbrace n_k \rbrace }e^{-i \int\lsz N_k(t) 
\rkop(0)+N_k^*(t)\rkdop(0)\rsz dk}\rhop_R(0)\ket{\lbrace n_k \rbrace}\nn ,
\end{align}
where $\rhop_S(0)$, $\rhop_{LB}(0)$, and $\rhop_R(0)$ characterize the initial 
coherences of the TLE, the link boson, and the reservoir, respectively.
The numbers $m$ and $ n_k $ represent the initial phonon number states of the 
LB and the reservoir mode $k$. 
Note that although here we only consider pure dephasing, a similar derivation can 
be given for other quantum noise effects, at least by using
established approximation schemes \cite{Weiss2012}.
In the following sections we discuss the results for examples (i) and (ii), introduced 
earlier in Section \ref{sec:model}, as well as in Fig.~\ref{fig:setup}(b) and (c), respectively.
\begin{figure}[b!]
\centering
\vglue -.4cm
\includegraphics[height=5.cm]{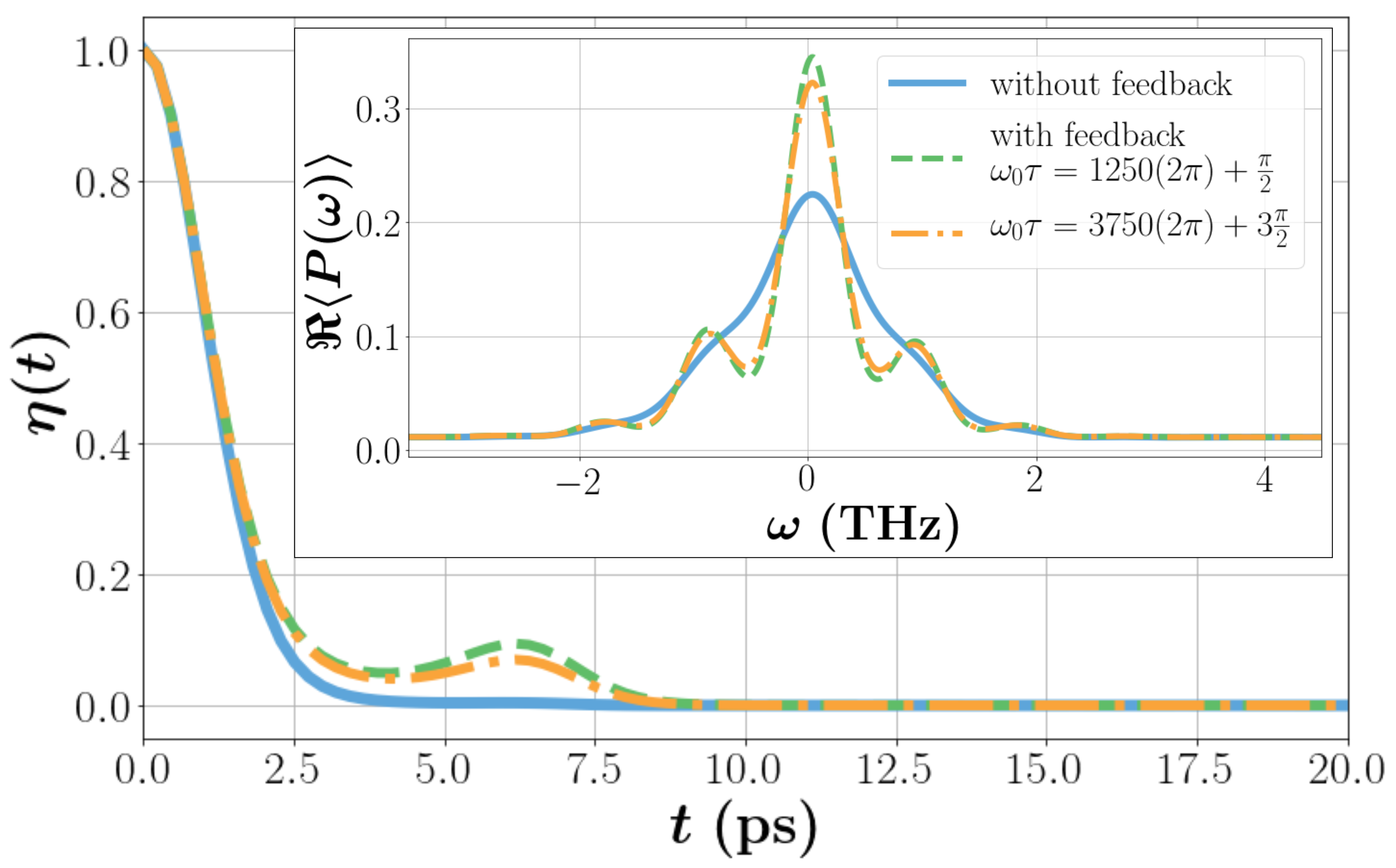}
\vglue -.3cm
\caption{The normalized absolute value of the TLE coherence, $\eta(t)$, as a function of time, with and without feedback, at 300~K. 
Inset: The corresponding absorption spectra. Parameters: 
$\omega_0=1$ THz, $D=200$ GHz, $\kappa=20$ GHz, $\kappa\tau=157.1\times\{1,3\}$ (with feedback).}
\label{fig:lw_narrow}
\end{figure}

\section{Example (i): LB exposed to Markovian loss}

Let us consider the reservoir coupling in the Markovian limit, $ g_k=g_0 $, and assume the same initial temperature for the reservoir
and the cavity. In order to include temperature, we use the canonical statistical operator $ 
\rho_R(0,T)=\exp[\sum_k\hbar\omega_k \rkdop(0)\rkop(0)/(k_BT)]/Z$,
giving the following contributions,
\begin{align*}
\sigma_R(t) &= \exp{\lka-\frac{1}{2}\int |N_k(t)|^2\lsz 2n_k(T)+1\rsz dk\rka},\\
\sigma_b(t) &= \exp{\lka-\frac{1}{2}|\gamma(t)|^2\lsz 2n_b(T)+1\rsz\rka},
\end{align*}
where $n_k(T)$ and $n_b(T)$ are the average occupation of the reservoir and 
LB modes at temperature $T$, respectively,
\begin{align*}
n_b&=\frac{1}{e^{\frac{\hbar\omega_0}{kT}}-1}&
n_k&=\frac{1}{e^{\frac{\hbar\omega_k}{kT}}-1}\\
\end{align*}
\vglue -.6cm
The blue, solid curves in Fig.~\ref{fig:lw_narrow} show the corresponding time trace $ \eta(t)$
and its Fourier transform (inset). The latter is proportional to the absorption
spectrum $ \mathcal{R}[\ew{\hat P(\omega)}]$, and is shown for parameters 
of a high-$\mathcal{Q}$ phonon cavity, in the regime of realizable
experimental platforms on nanofabricated hybrid-systems 
\cite{De2016,Rozas2009,Fainstein2013}.
%

Due to the Markovian reservoir, the coherence decays monotonically in proportion to 
the coupling strength $g_0^2$. 
The coherence is irreversibly lost to the reservoir modes
for longer times ($ \kappa t\gg1$) and, 
for the case shown in Fig.~\ref{fig:lw_narrow}, LB side-peaks or ``satellites'' in the absorption spectrum are broadened and barely noticeable due to 
the dissipation produced by the structureless reservoir, as expected 
\cite{Chernyak1996,Knox2002}. 
\subsection{Recovering the independent boson model}
A special case of the Markovian reservoir calculations is when we consider the temperature limit of $T=0$ for the reservoir, while keeping a finite temperature for the link boson. In this case, one finds
\begin{widetext}
\vspace{-.2cm}
\begin{align}
\label{eq:Pdamp_analytic}
&\Trace{\rhop^d_P(t)}=\frac{1}{2}\exp\left\{-\,\Theta(t)\frac{|D|^2}{|B|^2}\left[2B^*t+n_b\left(1+e^{
-2\kappa t}\right)+\frac{1}{2}\left(3-e^{-2\kappa 
t}\right)+i\frac{\omega_0^2}{|B|^2}\sin(\omega_0t)e^{-\kappa 
t}-\right.\right.\\
&\hspace{5.2cm}\left.\left.-\left(2n_b+1\right)\cos(\omega_0t)e^{-\kappa t}+\frac{2\kappa}{|B|^2}\left(-2B^*+B^*e^{-Bt}+Be^
{-B^*t}\right)\rsz+\right.\nn\\
\label{eq:shift}
&\hspace{3.2cm}\left.+i\,\Theta(t)\frac{|D|^2}{|B|^2}\lsz\lka 3\kappa\sin(\omega_0t)+4\omega_0\cos(\omega_0t)\rka\frac{\kappa}{|B|^2}e^{-\kappa 
t}-\frac{\omega_0}{2\kappa}\left(1-e^{-2\kappa 
t}\right)\right]\right\} .
\end{align}
\end{widetext}
The last row of this formula (\ref{eq:shift}) shows a shift of the satellites in the absorption spectrum compared to the expected values. These can become substantial if the couplings between the link boson and the emitter ($D$) or the reservoir ($g_0$) are comparable to the link boson frequency ($\omega_0$).

Without particle exchange with the environment, considering only pure dephasing, further simplification is possible. In this case the LB is not interacting with the reservoir at all $(g_0,\kappa\rightarrow0)$ and thus we arrive at the formula
\begin{widetext}
\vspace{-.2cm}
\begin{align}
\label{eq:rhop_ibm}
\rhop_P^\text{ibm}(t) =&\ \sigma_e(0) \exp\lka 
i\frac{D^2}{\omega_0}t-\frac{D^2}{\omega_0^2}\lsz(1+n_b)(1-e^{-i\omega_0 
t})+n_b\lk1-e^{i\omega_0 t}\rk\rsz\rka,
\end{align}
\end{widetext}
which is the result known from the independent boson model \cite{Nazir2015,Krummheuer2002}. 
\section{Example (ii): LB exposed to non-Markovian quantum feedback}
The dynamics changes significantly when the link boson couples to 
a structured reservoir.
The strength of the approach developed in this work is that we can treat 
such a reservoir with finite occupation and with an assumed boundary 
condition (here at $ L$ \cite{Krummheuer2005}). The distant, perfectly reflecting mirror depicted in Fig.~\ref{fig:setup}(c) introduces 
coherent feedback into the S-LB dynamics, as well as an entangled 
reservoir-LB dynamics \cite{Faulstich2017}.

\begin{figure}[b!]
\centering
\vglue -.4cm
\includegraphics[height=5.cm]{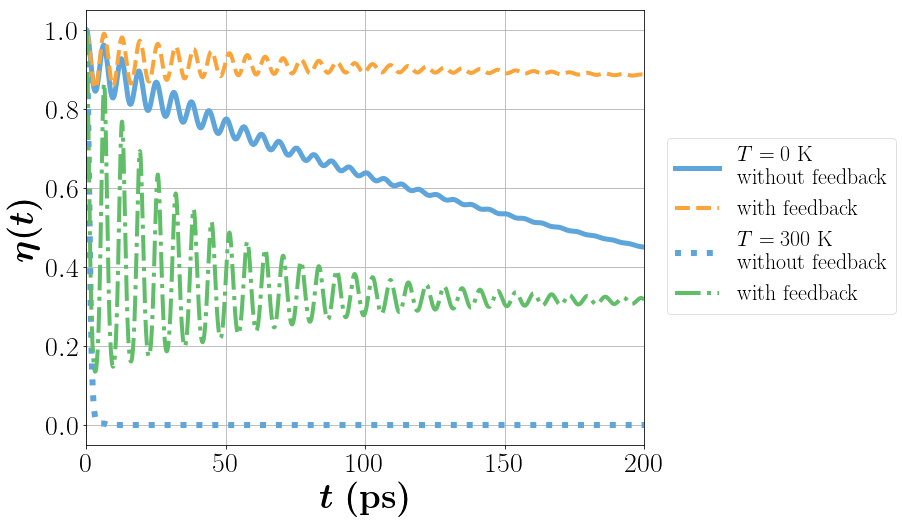}\\
\vglue -.3cm
\caption{Time-evolution of the normalized coherence, $\eta(t)$, with and without feedback, and for 
different temperatures. Parameters: $\omega_0=1$ THz, $D=200$ GHz, $\kappa=20$ 
GHz, $\kappa\tau=0.031$.
} 
\label{fig:temp_dep}
\end{figure}

This coherent feedback \cite{Zhang2017a}, originally introduced as an all-optical feedback \cite{Wiseman1994} for quantum systems, has been proven an
efficient way to recover lost quantum information from a reservoir 
with infinite degrees of freedom. Besides introducing a distant mirror, as just described \cite{Faulstich2017}, it can also be considered as a special case of cascaded quantum systems, as illustrated in Fig. \ref{fig:setup}(d) and described in \cite{Whalen2017}

For optical excitations, the effect of feedback first manifests itself as a reduction in the effective cavity
linewidth. In certain regimes, the time-delay associated with the feedback propagation also becomes an important dynamical control parameter. This was used to enhance intrinsic quantum properties of various systems, such as squeezing
\cite{Gough2009,Iida2011,Kraft2016,Nemet2016}, or to recover Rabi 
oscillations \cite{Kabuss2015}.
It has also proven to be useful for the manipulation 
of steady state behaviour of a given quantum system 
\cite{Nemet2016,Grimsmo2014} and to prepare various quantum states \cite{Kashiwamura2018}.
Here, for the LB-feedback case, we examine two limiting 
cases: the long-delay (Fig.~\ref{fig:lw_narrow}) and short-delay (Fig.~\ref{fig:temp_dep}) limits.
In the long-delay limit ($ \kappa\tau\gg1$), the feedback loop reduces the decay
rate of the coherence. 
In Fig.~\ref{fig:lw_narrow} (dashed and dashed-dotted lines) 
the LB-TLE interference is restored in the time-domain due to the 
feedback mechanism. This appears as a reduced effective linewidth
of the LB-satellites in the absorption spectrum.
However, we see that the feedback phase, i.e., the specific position of the 
reflecting surface $(\omega_0\tau=2k_0L)$ has only a weak impact; in particular, the 
green (dashed) and orange (dashed-dotted) lines in Fig.~\ref{fig:lw_narrow}
are almost identical. 
The only difference arises from the fact that for
decreasing delay, it is more probable that the TLE and the LB interact, as the cavity is more likely to be excited at a given point in time.
However, generally, in the long-delay limit $(\kappa\tau\gg 1)$, the LB excitation is absorbed
entirely by the reservoir before being fed back.
Thus, the significance of the specific phase $\omega_0\tau$ is negligible and, 
in fact, eventually the whole coherence is lost into the reservoir regardless of 
the reduced, effective linewidth.
\begin{figure}[t!]
\centering
\includegraphics[height=5.3cm]{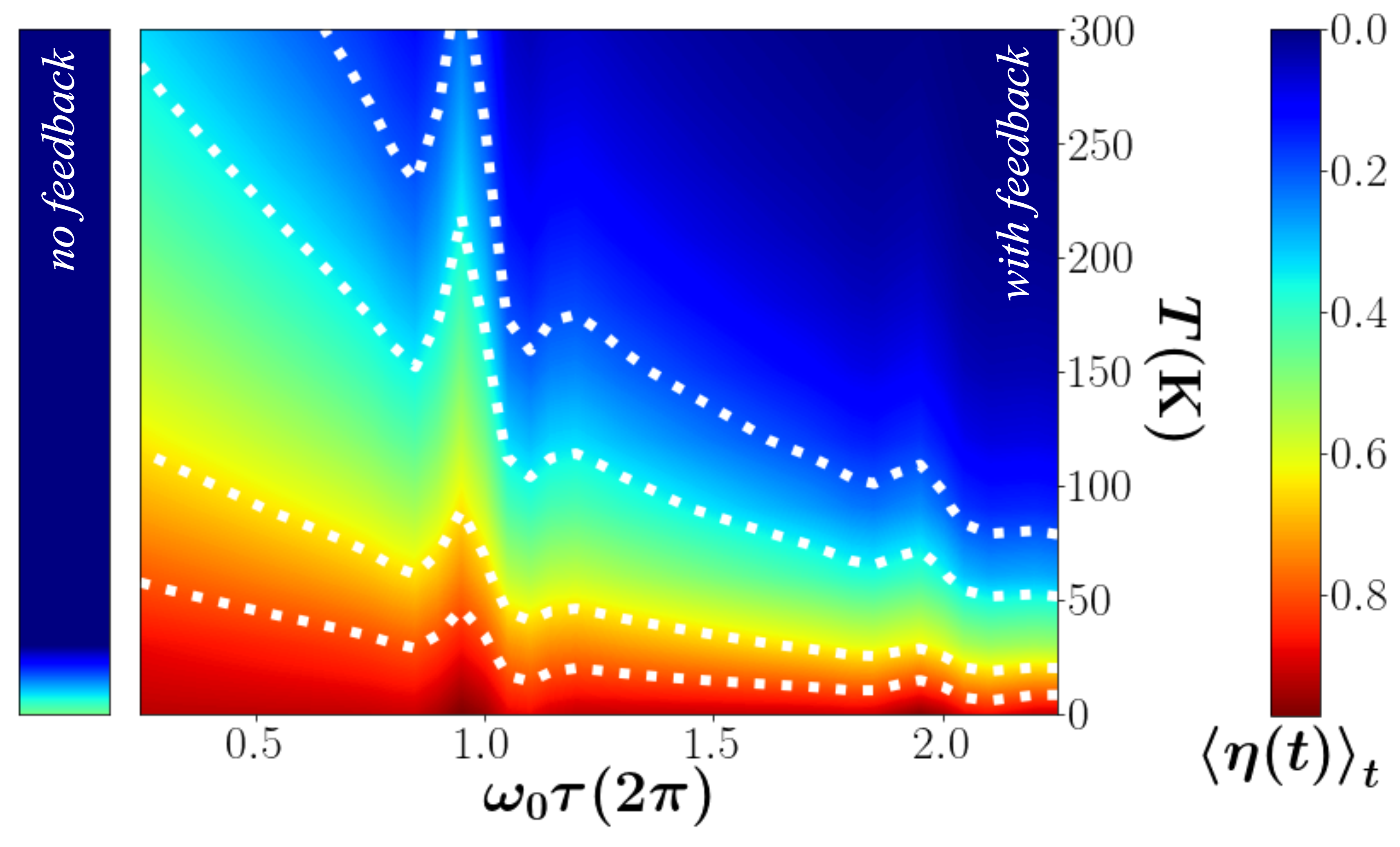}
\vglue -.3cm
\caption{Colormap of the normalized coherence, $\eta(t)$, at $t=200$~ps. {\it Without feedback}
the system decoheres rapidly at higher temperatures, whereas {\it with feedback} a 
finite coherence is preserved even at room temperature. Parameters: 
$\omega_0=1$ THz, $D=200$ GHz, $\kappa=20$ GHz. 
}
\label{fig:colormap}
\vglue -.4cm
\end{figure}

On the other hand, in the case of short delay ($ \kappa\tau\ll1 $), quantum 
interferences occur between the LB and the reservoir. This is because, with decreasing 
time delay $ \tau $, there is a higher chance 
that the feedback signal observes a finite LB-excitation. 
This results in an oscillating coherence of the TLE, which shows the recovered coherent dynamics of the LB and the system;  
see Fig.~\ref{fig:temp_dep} (dashed and dash-dotted lines). 
The introduced memory of the 
environment preserves a large portion of the initial coherence of the TLE,
in contrast to the case without feedback, when all coherence is 
inevitably lost. This effect arises even for an initial thermal state of the LB and  reservoir; see Fig.~\ref{fig:temp_dep}.  
Note that, however, the amount of leftover coherence decreases with growing temperature of
the phonon reservoir, as shown in Fig.~\ref{fig:colormap} (see also  
\cite{Nazir2015}), although, interestingly, persistence of the initial coherence can hold up to very high
temperatures, in contrast to the case without feedback, where the coherence
is damped even at $T=0$ due to spontaneous emission of the LB into its 
reservoir.
\vspace{-.3cm}
\subsection{Influence of the feedback phase}

\begin{figure}[t!]
\centering
\includegraphics[width=8.3cm]{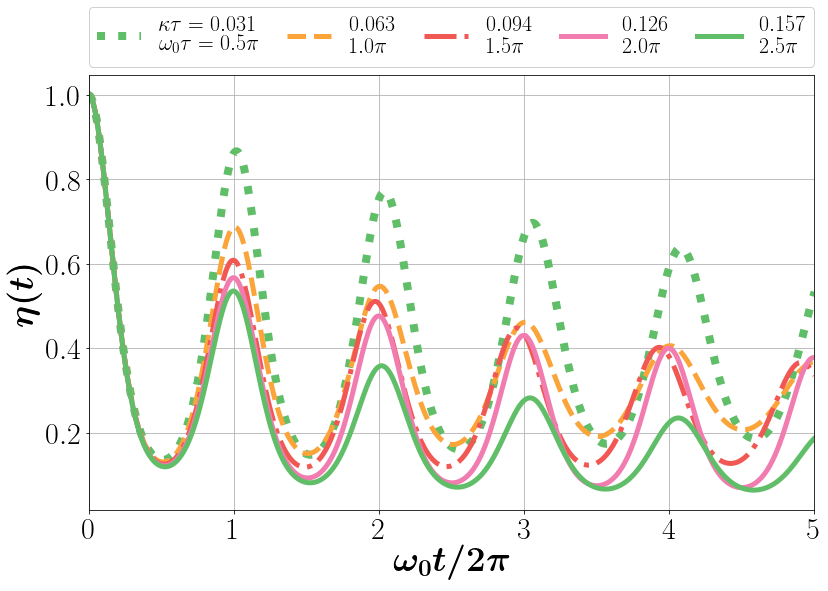}\\
\vglue -.3cm
\caption{Time-evolution of the coherence with feedback for different 
time-delays. Parameters: $T=300$ K, $\omega_0=1$ THz, $D=200$ GHz, $\kappa=20$ 
GHz. }
\label{fig:phase_diff}
\end{figure}

In the case of time-delayed coherent feedback, a short feedback length means 
that the accumulated propagation phase ($\omega_0\tau$) becomes an important control parameter. At phases described by $n\pi$, where $n\in\{1,2,...\}$ we recover oscillations with the LB frequency ($\omega_0$), whereas the 
frequency slightly changes for other phases, as can be observed in Fig.~\ref{fig:phase_diff} for feedback phases $(n-1/2)\pi$.

\begin{figure}[t!]
\centering
\includegraphics[height=5.3cm]{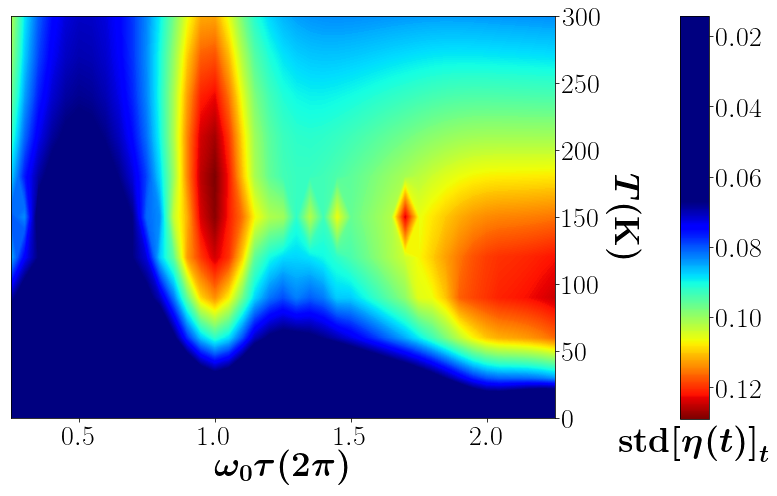}\\
\vglue -.3cm
\caption{The deviation of the normalized coherence, $\eta(t)$,  from its mean value over the time period of 200 ps 
as a function of temperature and feedback phase. In the case of constructive interference (integer multiples of $2\pi$),
more pronounced oscillations can be observed, which appear as "hot spots" in the figure. On the other hand, for destructive interference (odd integer multiples of $\pi$), 
the amplitudes of the recovered oscillations are smaller, which appear as a smaller deviation from the mean value. Parameters: $T=300$ K, $\omega_0=1$ THz, $D=200$ GHz, $\kappa=20$ 
GHz. }
\label{fig:colorstd}
\vglue -.2cm
\end{figure}

Note that the feedback phase has a less significant effect on the amount of leftover 
coherence than the length of the corresponding time-delay. Its contribution becomes
visible mainly in the frequency of oscillations and in the amplitude of recovered
oscillations. The greater these oscillations are, the larger is the deviation from the mean value in the normalized coherence over the time period shown in Fig.~\ref{fig:temp_dep}. Thus we calculated the standard deviation of $\eta(t)$ as in eq.(\ref{eq:eta}) over the first 200 ps as a function of temperature and 
propagation phase (Fig.~\ref{fig:colorstd}).

Note that for lower temperatures, even though the steady state coherence is substantial (Fig.~\ref{fig:temp_dep}), the amplitude of the oscillations is quite small. This also shows up in Fig.~\ref{fig:colorstd}, where signs of significant oscillations can only be seen for temperatures above $50$ K. The "hot spots" coincide well with the propagation phases, where a peak can be observed in the recovered coherence in Fig.~\ref{fig:colormap}, i.e., at multiples of $2\pi$. 

\section{Conclusion}
We have given a general framework for the description of structured-reservoir-induced quantum dynamics via the introduction of an additional link boson interaction, $\hat H_{LB}$, to the system 
Hamiltonian $\hat H_S$.
In the case of a TLE coupled to the LB, an analytical expression is obtained 
for the resulting unconventional dephasing dynamics.
Due to the analytical nature of the solutions, this method is readily capable of treating arbitrary initial states for both the LB and the reservoir.  For example, one may consider initial Fock-state excitation of the LB ( see Appendix \ref{sec:Fock}).

Our investigations demonstrate a robust recovery of LB-system 
oscillations in a continuous environment with structured coupling. In the short-delay limit the propagation phase in the feedback loop $(\omega_0\tau)$ becomes an important control parameter, influencing both the frequency and amplitude of the recovered oscillations. 

We also show that using this specific interaction between the LB and the environment, the initial coherence of the electronic system can be stabilized even at finite temperatures. 
Thus the introduced theoretical description opens up a new horizon for reservoir manipulation as, to the best of our knowledge, calculations for a quantum system with time-delayed coherent feedback at a finite temperature have not been performed before.

\section*{Acknowledgements.}
NN, AK, and AC gratefully acknowledge the support of the
Deutsche Forschungsgemeinschaft (DFG) through the project B1 of the SFB 910 and 
by the school of nanophotonics (SFB 787), and discussions with A. Metelmann.
A.K. thanks the Auckland group for the hospitality, and A.C. thanks J. 
Specht, V. Dehn and J. Kabuss for fruitful discussions. N.N. thanks the group
of A.K. in Berlin for the support and hospitality.

The authors would also like to thank the anonymous Referees for their insights, which greatly improved the overall quality of the manuscript.
%
\appendix
\section{Derivation of the coefficients in the linear map with feedback \label{sec:FG}}

The main point of taking the interaction picture is to describe the environment's influence
on the system only through its coupling to the LB by introducing an effective time dependence.
The interaction picture Hamiltonian is considered as
\begin{align}
\label{eq:HI}
&\Hop_I/\hbar = \Hop_R/\hbar+\Hop_S/\hbar=\\
&\hspace{.4cm}=\Hop_S/\hbar+ 
\omega_0 \bdop\bop\ +  
\int \left[  \omega_k \rkdop\rkop  + 
g_k(\rkdop\bop+\bdop\rkop) \right]dk,\nonumber
\end{align}
from which the following equations of motion can be derived for the reservoir operators
and the LB:
\begin{align*}
\dot{\bop} &= -i\omega_0\bop(t)-i\int g_k\rop_k(t)dk,\\
\dot{\rop}_k &= -i\omega_k\rop_k(t)-ig_k\bop(t).
\end{align*}
Note that, by changing $g_k$ to $g_0$, we obtain the usual Markovian time-evolution.
By formally integrating the second equation and substituting it back into the 
equation of motion for the LB, we obtain the following:
\begin{align*}
\dot{\bop} =& -i\omega_0\bop(t) -i\int  g_ke^{-i\omega_k t}\rop_k(t) dk -\nn\\
&-\int_0^t\bop(t^\prime)\int  g_k^2e^{-i\omega_k(t-t^\prime)}dkdt^\prime .
\end{align*}
The last term describes the environmental back-action on the state of the link-boson. Introducing a specific boundary condition for the problem results in a coherent feedback
$\lk g_k = g_0\sin{(kL)}=g_0\sin{\lk\frac{\omega_k\tau}{2}\rk}\rk$ and a spectral density \cite{Petruccione2007}
\begin{align}
&J(\omega_k) = \sin^2{\left(\frac{\omega_k\tau}{2}\right)}e^{-i\omega_k(t-t^\prime)}=\\
&\hspace{.2cm}=\frac{1}{2}\left(e^{-i\omega_k(t-t^\prime)}+\frac{1}{2}e^{-i\omega_k(t-t^\prime+\tau)}+\frac{1}{2}e^{-i\omega_k(t-t^\prime-\tau)}\right),\nn
\end{align}
where the first term describes the standard Markovian decay, the second refers to a non-causal dependence, and the last term describes the past state of the link boson field fed back by the environment. Integrating over the frequencies and taking causality into account, we obtain the following for the environmental back action,
\begin{widetext}
\vspace{-.5cm}
\begin{align*}
-\int_0^t\bop(t^\prime)\int g_0^2 J(\omega_k) dkdt^\prime &
= \frac{g_0^2\pi}{2c}\int_0^t\bop(t^\prime)\lsz\delta(t-\tp+\tau)+\delta(t-\tp-\tau)-2\delta(t-\tp)\rsz dt=\kappa\lsz\bop(t-\tau)-\bop(t)\rsz .
\end{align*}
\vspace{-.2cm}
\end{widetext}
Thus, the equation of motion for the LB with the effective action of the environment becomes
\begin{align}
\label{eq:LBeqmot}
\dot{\bop} =& -B\bop(t)-i\int g_ke^{A_kt}\rop_k(0)dk+\nn\\
&+\kappa\Theta(t-\tau)\bop(t-\tau),
\end{align}
where we use
\begin{align*}
B&=i\omega_0+\kappa,\hspace{1cm} A_k = -i\omega_k,\hspace{1cm}\kappa = \frac{\pi g_0^2}{2c}.
\end{align*}

The usual way of dealing with time-delayed dynamic equations is to apply the method of steps, where the solution is evaluated
for each $\tau$-interval by first ignoring the delay-term and then substituting the solution obtained for the previous $\tau$-interval into the last term.

Let us follow this through the example of the first two $\tau$-intervals. For $0<t<\tau$ we have the following equation of motion,
\begin{align*}
\dot{\bop} &= -B \bop(t) +\int D_k(t)\rop_k(0) dk,\\
D_k(t)& = -ig_ke^{A_k t},
\end{align*}
for which we can obtain the solution
\begin{align*}
\bop(t) &= e^{-B t}\bop(0)+\int G_0^k(t)\rop_k(0)dk,\\
G_0^k(t) &= \int_0^t D_k(t^\prime)e^{-B(t-t^\prime)}d\tp=-\frac{ig_k}{A_k+B}\lk e^{A_kt}-e^{-Bt}\rk.
\end{align*}
By exchanging $g_k$ for the $k$-independent $g_0$, this recovers the Markovian coefficient 
$G^d_k(t)$ as in Eq.~(5) in the main text.
Substituting the above solution back into the last term of Eq.~(\ref{eq:LBeqmot}) gives the following
equation of motion for the time-interval $0<t<2\tau$,
\begin{align*}
\dot{\bop} =& -B\bop(t)+\kappa\Theta(t-\tau)\bop(t-\tau)+\nn\\
&+\int\lsz D_k(t)\kappa\Theta(t-\tau)G_0^k(t-\tau)\rsz \rop_k(0)dk,
\end{align*}
for which we can obtain the solution
\begin{align*}
\bop(t) &= \lsz \Theta(t)e^{-Bt}+\kappa\Theta(t-\tau)e^{-B(t-\tau)}(t-\tau)\rsz \bop(0) + \nn\\
&\hspace{.5cm}+\int\lsz\Theta(t)G_0^k(t)+\Theta(t-\tau)G_1^k(t)\rsz \rop_k(0)dk,
\end{align*}
\begin{align*}
G_1^k(t) &= \kappa\int^t_\tau G_0^k(\tp-\tau)e^{-B(t-\tp)}d\tp=\nn\\
&=-\frac{ig_k\kappa}{(A_k+B)^2}\lka e^{A_k(t-\tau)}-\right.\nn\\
&\left.\hspace{2.5cm}-e^{-B(t-\tau)}\lsz 1+(A_k+B)(t-\tau)\rsz\rka.
\end{align*}

Continuing in a similar fashion with the other time-intervals, the following ansatz can be assumed
for a general coefficient $G_m^k(t)$:
\begin{align*}
G_m^k(t) =& -\frac{ig_k\kappa^m}{(A_k+B)^{m+1}}\lsz e^{A(t-m\tau)}-\right.\nn\\
&\left.\hspace{1cm}-e^{-B(t-m\tau)}\sum_{n=0}^m\frac{(t-m\tau)^n}{n!}(A_k+B)^n\rsz.
\end{align*}

The next step in our derivation is to prove that by using the general recursive method
demonstrated for $G_1^k(t)$, we obtain the next coefficient in the same form ($M=m+1$):
\begin{widetext}
\vspace{-.5cm}
\begin{align*}
G_{M}^k(t) &= \kappa\int_{(m+1)\tau}^tG_m^k(\tp-\tau)e^{-B(t-\tp)}d\tp = \\
&=-\frac{ig_k\kappa^Me^{-Bt}}{(A_k+B)^M}\lsz\int_{M\tau}^te^{(A_k+B)\lk\tp-M\tau\rk}e^{BM\tau}dt^\prime-e^{BM\tau}\sum_{n=0}^m\frac{(A_k+B)^n}{n!}\int_{M\tau}^t\lk\tp-M\tau\rk^nd\tp\rsz=\\
&=-\frac{ig_k\kappa^Me^{-B(t-M\tau)}}{(A_k+B)^M}\lsz\frac{e^{(A_k+B)(t-M\tau)}-1}{A_k+B}-\sum_{n=0}^m\frac{(A_k+B)^n}{n!}\frac{(t-M\tau)^{n+1}}{n+1}\rsz=\\
& = -\frac{ig_k\kappa^M}{(A_k+B)^{M+1}}\lsz e^{A(t-M\tau)}-e^{-B(t-M\tau)}\sum_{n=0}^M\frac{(t-M\tau)^n}{n!}(A_k+B)^n\rsz.
\end{align*}
\end{widetext}

For the coefficients of $\bop(0)$ a similar process leads to the general formula
\begin{align*}
F_m(t) = \frac{\kappa^m}{m!}e^{-B(t-m\tau)}(t-m\tau)^m.
\end{align*}
By summing up these contributions as
\begin{align*}
F_{fb}(t) = \sum_{m=0}^\infty \Theta(t-m\tau)F_m(t),\\
G_k^{fb}(t) = \sum_{m=0}^\infty\Theta(t-m\tau) G_m^k(t),
\end{align*}
we obtain the final form of the coefficients given in the main text in Eqs.~(6,7).

\section{Alternative derivation for a pure dephasing interaction\label{sec:sp-LB}}
Since time-dependence of $ \hat H_{LB}(t)$  can be 
expressed solely by $c$-number functions and initial operators that are time-independent, (Eq.~(\ref{eq:FMark}-\ref{eq:Gfb})), an exact formula can 
be obtained for the dynamics
of the observable $ \hat P(t) $: 
\begin{align}
i\hbar \dt{\rhop_I}\Pop(t) &= \lsz\Hop_{LB}, \rhop_I\rsz\Pop(t),\nn\\
\dt{\rhop_I}\Pop(t) &=-iD\lsz \bop(t)+\bop^\dg(t) \rsz \Pop^\dg(t) \Pop(t) 
\rhop_I\Pop(t).
\end{align}
By exploiting the intrinsic linear dynamics of the TLE coherence 
(\ref{eq:P_evol}) we obtain 
\begin{align}
\dt{\rhop_P}&=-i \Bop(t)  \Pop^\dg(0) \Pop(0) \rhop_P, \\
\Bop(t) &= \bop^\dg(t) + \bop(t),
\end{align}
where we abbreviate $\hat \rho^I(t) \hat P(t)=:\hat \rho^I_P(t)$, and 
$ \text{Tr}[\rhop^I_{\hat P}(t)]=\ew{\hat P(t)} $.
A stroboscopic solution of the equation of motion for short enough time steps gives
\begin{align*}
&\rhop_P(t+\Delta t)=\lk 1-i \Bop(t) \Pop^\dg(0) \Pop(0) \Delta t\rk \rhop_P(t) \approx
\\
&\approx e^{-i \Bop(t)\Pop^\dg(0) \Pop(0) \Delta t}\, e^{-i \Bop(t-\Delta t) \Pop^\dg(0) \Pop(0) \Delta t}\rhop_P(t-\Delta t).
\end{align*}
By applying the Baker-Campbell-Hausdorff formula, this translates into
\begin{align}
&\hat \rho^I_{\Pop}(t) =
\exp
\left[
-i
\int_0^t 
\Hop_{LB}(t_1) 
dt_1
\right]\times\\  \nonumber
&\hspace{.6cm} \times\exp
\left[
-\frac{1}{2}
\int_0^t 
\int_0^{t_1} [\Hop_{LB}(t_1),\Hop_{LB}(t_2)]dt_2
dt_1
\right]
\hat
\rho^I_{\Pop}(0),
\end{align}
which is the same as was obtained in (\ref{eq:20}).

\section{Initial Fock excitation in the LB\label{sec:Fock}}

\noindent Let us assume $n$ bosons for the LB in the beginning 
$\lk\sigma_{LB}(0)=\ketbra{n}{n}\rk$ and 0 K for the reservoir. Then, the link boson part of the expectation value (\ref{eq:sigma_def}) becomes
\begin{align}
\sigma_{LB}(t)  &=\sum_{m}\bra{m}e^{-i \lsz\gamma(t) 
\bop(0)+\gamma^*(t)\bop^\dg(0)\rsz}\ketbra{n}{n}\ket{m}=\nn\\
&=\bra{n}e^{-i \lsz\gamma(t) \bop(0)+\gamma^*(t)\bop^\dg(0)\rsz}\ket{n}.
\end{align}
If $n$ is set to $ 0 $, we have a series, which results in the typical exponential evolution,
\begin{align}
\sigma_{LB}(t)  =&\ \exp\lsz-\frac{|\gamma(t)|^2}{2} \rsz ,
\end{align}
which is equivalent to the zero temperature case. Higher Fock-state contributions can be
calculated by using the Baker-Campbell-Hausdorff formula, since 
\begin{align}
[-i\gamma^*(t) \bop^\dg(0),-i\gamma(t) \bop(0)] = |\gamma(t)|^2,
\end{align}
thus the link boson part of the expectation value becomes
\begin{align}
\sigma_{LB}(t)=& \bra{n}e^{-i \gamma^*(t)\bop^\dg(0)}e^{-i \gamma(t) 
\bop(0)}\ket{n}e^{-\frac{|\gamma(t)|^2}{2}},
\end{align}
which can be evaluated as
\begin{widetext}
\begin{align}
\sigma_{LB}(t)&=\lk\bra{n}e^{-i\gamma^*(t)\bop^\dagger(0)}\rk\lk 
e^{-i\gamma(t)\bop(0)}\ket{n}\rk e^{-\frac{|\gamma(t)|^2}{2}}=\nn\\
&=\lk\bra{n}\sum_{l=0}^n\frac{\lsz-i\gamma^*(t)\bop^\dagger(0)\rsz^l}{l!}
\rk\lk\sum_{l^\prime=0}^n\frac{\lsz-i\gamma(t)\bop(0)\rsz^{l^\prime}}{
l^\prime!}\ket{n}\rk e^{-\frac{|\gamma(t)|^2}{2}} =\nn\\
&=\sum_{l=0}^n\frac{\left[-|\gamma(t)|^2\right]^l}{l!}\binom{n}{l}e^{-\frac{
|\gamma(t)|^2}{2}}\nn.
\end{align}
\end{widetext}

\newpage

\bibstyle{apsrev4-1}
%
\end{document}